# PERFORMANCE ANALYSIS OF ULTRA WIDEBAND RECEIVERS FOR HIGH DATA RATE WIRELESS PERSONAL AREA NETWORK SYSTEM

Bikramaditya Das[*] and Susmita Das, Member, IEEE[*]

[*]Department of Electrical Engineering, National Institute of Technology,
Rourkela -769008, India
adibik09@gmail.com

sdas@nitrkl.ac.in

## ABSTRACT

*For high data rate ultra wideband communication system, performance comparison of Rake, MMSE and Rake-MMSE receivers is attempted in this paper. Further a detail study on Rake-MMSE time domain equalizers is carried out taking into account all the important parameters such as the effect of the number of Rake fingers and equalizer taps on the error rate performance. This receiver combats inter-symbol interference by taking advantages of both the Rake and equalizer structure. The bit error rate performances are investigated using MATLAB simulation on IEEE 802.15.3a defined UWB channel models. Simulation results show that the bit error rate probability of Rake-MMSE receiver is much better than Rake receiver and MMSE equalizer. Study on non-line of sight indoor channel models illustrates that bit error rate performance of Rake-MMSE (both LE and DFE) improves for CM3 model with smaller spread compared to CM4 channel model. It is indicated that for a MMSE equalizer operating at low to medium SNR values, the number of Rake fingers is the dominant factor to improve system performance, while at high SNR values the number of equalizer taps plays a more significant role in reducing the error rate.*

## KEYWORDS

*UWB, Rake receiver, MMSE, Rake-MMSE, Bit Error Rate, LE, DFE.*

## 1. INTRODUCTION

Ultra-wideband (UWB) has recently evoked great interest and its potential strength lies in its use of extremely wide transmission bandwidth. Furthermore, UWB is emerging as a solution for the IEEE 802.15a (TG3a) standard which is to provide a low complexity, low cost, low power consumption and high data-rate among Wireless Personal Area Network (WPAN) devices. An aspect of UWB transmission is to combat multipath propagation effects. Rake receivers can be employed since they are able to provide multipath diversity [1-3]. Another aspect is to eliminate or combat the inter-symbol interference (ISI) which distorts the transmitted signal and causes bit errors at the receiver, especially when the transmission data rate is very high as well as for which are not well synchronized. In [1] and [3], the "rake decorrelating effect" was mentioned as a way to combat ISI. Combination of spatial diversity combining and equalization is a well established scheme for frequency selective fading channels. In [5], a combined rake and equalizer structure was proposed for high data rate UWB systems. In this paper, the performance of a rake-MMSE-equalizer receiver similar to [5] is investigated for different number of rake fingers and equalizer taps using a semi-analytical approach. We propose at first to study time equalization with combined Rake-MMSE equalizer structure. We show that, for a MMSE equalizer operating at low to medium SNR's, the number of Rake fingers is the dominant factor to improve system performance, while, at high SNR's the number of equalizer taps plays a more significant role in reducing error rates[7-8]. We show that for high frequency selective channels such as the CM4 one, a linear equalizer structure is not sufficient and must be replaced by a decision feedback equalizer (DFE) structure. Furthermore, we propose a simple recursive gradient based algorithm to implement the equalizer structures.

The rest of the paper is organized as follows. In Section 2 we study the signals and system model for IEEE UWB channel modelling. Section 3 is devoted principles of equalizations and receiver structure. In section 4 we study performance analysis for Rake-MMSE receiver. Simulation results are discussed in Section 5. Section 6 concludes the paper.

## 2. SIGNALS AND SYSTEM MODEL

For a single user system, the continuous transmitted data stream is written

$$s(t) = \sum_{k=-\infty}^{+\infty} d(k).p(t - k.T_s) \qquad (1)$$

Where $d(k)$ are stationary uncorrelated BPSK data and $T_s$ is the symbol duration. Throughout this paper we consider the application of a root raised cosine (RRC) transmit filter $p(t)$ with roll-off factor $\alpha = 0.3$. The UWB pulse $p(t)$ has duration $T_{uwb}$ ($T_{uwb} < T_s$ ).

The channel models used in this paper are the model proposed by IEEE 802.15.3a Study Group [10]. In the normalized models provided by IEEE 802.15.3a Study Group, different channel characteristics are put together under four channel model scenarios having rms delay spreads ranging from 5 to 26 nsec. For this paper four channel models, derived from the IEEE 802.15 channel modelling working group. In IEEE 802.15.3a working group, the UWB channel is further classified into four models. Channel model 1 (CM1) represents LOS and distance from 0 to 4 m UWB channel, while channel model 2 (CM2) represents NLOS and distance from 0 to 4 m UWB channel. Distance from 4 m to 10 m and NLOS UWB channel is modelled as CM3 and distance over 10 m NLOS UWB channels are all classified into the extreme model CM4.

The impulse response can be written as

$$h(t) = \sum_{p=0}^{M} h_p.\delta(t - \tau_p) \qquad (2)$$

Parameter $M$ is the total number of paths in the channel.

## 3. PRINCIPLE OF RECEIVERS STRUCTURE

### 3.1. RAKE RECEIVER

Rake receivers are used in time-hopping impulse radio systems and direct sequence spread spectrum systems (DS-SS) for matched filtering of the received signal [9]. The receiver structure consists of a matched filter that is matched to the transmitted waveform that represents one symbol, and a tapped delay line that matches the channel impulse response. It is also possible to implement this structure as a number of correlators that are sampled at the delays related to specific number of multipath components; each of those correlators can be called "Rake finger." A Rake receiver structure is shown in Fig.1.

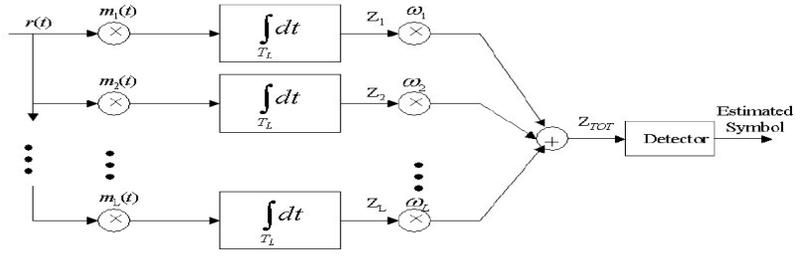

Figure.1. UWB RAKE receiver structure

The received signal first passes through the receiver filter matched to the transmitted pulse and is given by

$$r(t) = s(t) * h(t) * p(-t) + n(t) * p(-t)$$

$$= \sum_{k=-\infty}^{+\infty} d(k) \sum_i h_i . m(t - k.T_s - \tau_i) + \overset{\wedge}{n}(t) \qquad (3)$$

Where $p(-t)$ represents the receiver matched filter, "*" stands for convolution operation and $n(t)$ is the additive white Gaussian noise (AWGN) with zero mean and variance $N_0/2$. Also, $m(t) = p(t) * p(-t)$ and $n(t) = n(t) * p(-t)$.

Combining the channel impulse response (CIR) with the transmitter pulse shape and the matched filter, we have

$$\tilde{h}(t) = p(t) * h(t) * p(-t) = \sum_{i=0}^{M} h_i . m(t - \tau_i) \qquad (4)$$

The output of the receiver filter is sampled at each Rake finger. The minimum Rake finger separation is $T_m = T_s / N_u$, where $N_u$ is chosen as the largest integer value that would result in $T_m$ spaced uncorrelated noise samples at the Rake fingers

$$v(nT_s + \tau'_i + t_0) = \sum_{k=-\infty}^{+\infty} \tilde{h}((n-k)T_s + \tau'_i + t_0)d(k) \qquad (5)$$

where $\tau_l'$ is the delay time corresponding to the $l^{th}$ Rake finger and is an integer multiple of $T_m$. Parameter $t_0$ corresponds to a time offset and is used to obtain the best sampling time. Without loss of generality, $t_0$ will be set to zero in the following analysis.

## 3.2. MMSE STRUCTURE

In reality the noise component due to the physical channel cannot be ignored. In the presence of additive Gaussian noise at the receiver input, the output of the equalizer at the $n^{th}$ sampling instant is given by

$$\hat{y}_n = \sum_{K=-N}^{N} b_k r_{n-k} \qquad (6)$$

The mean square error (MSE) for the equalizer having $2N+1$ taps, denoted by $J(N)$ is

$$J(N) = E|x_n - \hat{y}_n|^2 = E\left[\left(x_n - \sum_{k=-N}^{N} b_k r_{n-k}\right)^2\right] \qquad (7)$$

$J(N)$ with respect to the equalizer coefficients $(b_k)$ is obtained by the following differentiation:

$$\frac{dJ(N)}{db_k} = 0 \qquad (8)$$

Equation (7) leads to the necessary condition for the minimum MSE given by

$$R_r b = R_{xr} \qquad b = R_r^{-1} R_{xr} \qquad (9)$$

WHERE

$$R_{xr} = (R_{xr}(-N)....R_{xr}(0)....R_{xr}(N))^T \qquad (10)$$

$$R_r = \begin{bmatrix} R_r(0)....... & R_r(N)....... & R_r(2N) \\ . & . & . \\ . & . & . \\ R_r(-2N+1) & R_r(-N+1) & R_r(1) \\ R_r(-2N)_{2N} & R_r(-N) & R_r(0) \end{bmatrix} \qquad (11)$$

## 3.3. RAKE-MMSE STRUCTURE

The receiver structure is illustrated in Fig. 2 and consists in a Rake receiver followed by a linear equalizer. As we will see later on, a structure gives better performances over UWB channels when the number of equalizer taps is sufficiently large. The received signal first passes through the receiver filter matched to the transmitted pulse (3). The output of the receiver filter is sampled at each Rake finger. The minimum Rake finger separation is $T_m = T_s / N_u$ , where $N_u$ is chosen as the largest integer value that would result in Tm spaced uncorrelated noise samples at the Rake fingers(4). In a first approach, complete channel state information (CSI) is assumed to be available at the receiver. For general selection combining, the Rake fingers ($\beta$'s) are selected as the largest $L$ ($L \leq N_u$) sampled signal at the matched filter output within one symbol time period at time instants $\tau_l'$ ,$l = 1, 2, ..., L$ . A feasible implementation of multipath diversity combining can be obtained by a selective-Rake (SRake) receiver, which combines the $L$ best, out of $N_u$, multipath components. Those L best components are determined by a finger selection algorithm. For a maximal ratio combining (MRC) Rake receiver, the paths with highest signal-to-noise ratios (SNRs) are selected, which is an optimal scheme in the absence of interfering users and intersymbol interference (ISI). For a minimum mean square error (MMSE) Rake receiver, the "conventional" finger selection algorithm is to choose the paths with highest signal-to-interference-plus-noise ratios (SINRs) [2]. The received signal sampled at the $l$ th Rake finger in the $n$th data symbol interval is given by equation (5).

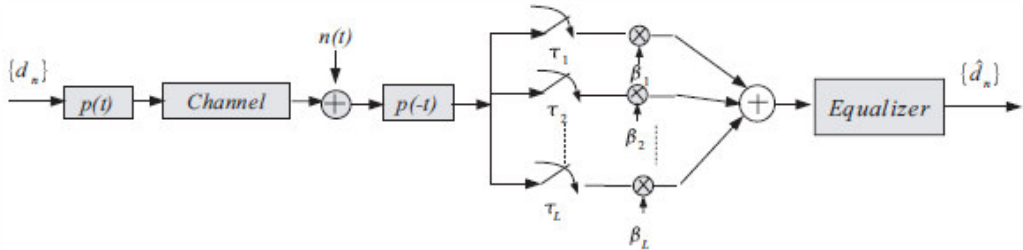

Figure.2. UWB RAKE-MMSE receiver structure

The Rake combiner output at time $t = n.T_s$ is

$$y[n] = \sum_{l=1}^{L} \beta_l . v(n.T_s + \tau_l') + \sum_{l=1}^{L} \beta_l . \overset{\wedge}{n}(n.T_s + \tau_l') \qquad (12)$$

Choosing the correct Rake finger placement leads to the reduction of ISI and the performance can be dramatically improved when using an equalizer to combat the remaining ISI. Considering the necessary tradeoff between complexity and performance, a sub-optimum classical criterion for updating the equalizer taps is the MMSE criterion.

## 4. PERFORMANCE ANALYSIS

In this part, due to the lack of place we will only discuss the matrix block computation of linear equalizers. Furthermore, we suppose perfect channel state information (CSI). Assuming that the n data bit is being detected, the MMSE criterion consists in minimizing

$$E\left[\left|d(n) - \hat{d}(n)\right|^2\right] \tag{13}$$

where $d(n)$ is the equalizer output. Rewriting the Rake output signal, one can distinguish the desired signal, the undesired ISI and the noise as

$$y(n) = \left[\sum_{l=1}^{L}\beta_l.\tilde{h}(\tau_l)\right].d(n) + \sum_{k \neq n}\sum_{l=1}^{L}\beta_l.\tilde{h}\left((n-k)T_s + \tau_l\right)d(k) \tag{14}$$
$$+ \sum_{l=1}^{L}\beta_l.\hat{n}\left(n.T_s + \tau_l\right)$$

where the first term is the desired output. The noise samples at different fingers, $n(n.Ts + \tau_l')$, $l = 1... L$, are uncorrelated and therefore independent, since the samples are taken at approximately the multiples of the inverse of the matched filter bandwidth. It is assumed that the channel has a length of $(n_1 + n_2 + 1).T_s$. That is, there is pre-cursor ISI from the subsequent $n_1$ symbols and post-cursor ISI from the previous $n_2$ symbols, and $n_1$ and $n_2$ are chosen large enough to include the majority of the ISI effect. Using (8), the Rake output can be expressed now in a simple form as

$$y(n) = \alpha_0.d(n) + \sum_{\substack{k=-n_1 \\ k \neq 0}}^{n_2}\alpha_k.d(n-k) + \tilde{n}(n) \tag{15}$$

$$= \phi^T d[n] + \tilde{n}(n)$$

where coefficient $\alpha_K$'s are obtained by matching (14) and (15). $\phi = [\alpha_{n_1}...\alpha_0...\alpha_{n_2}]$ and $d[n] = [d(n+n_1)....d(n). ...d(n - n_2)]^T$ The superscript denotes the transpose operation. The output of the linear equalizer is obtained as

$$\hat{d}(n) = \sum_{r=-k_1}^{k_2} c_r.y(n-r) = c^T\gamma(n) + c^T\eta(n) \tag{16}$$

where $c = [c_{-K}...c_0...c_{K_2}]$ contains the equalizer taps. Also

$$\gamma[n] = \left[\phi^T d[n+K_1]..\phi^T d[n]..\phi^T d[n-K_2]\right]^T$$
$$\eta[n] = \left[\tilde{n}(n+K_1)...\tilde{n}(n)...\tilde{n}(n=K_2)\right]^T \tag{17}$$

The mean square error (MSE) of the equalizer,

$$E\left[\left|d(n) - c^T\gamma[n] - c^T\eta[n]\right|^2\right] \tag{18}$$

which is a quadratic function of the vector $c$, has a unique minimum solution. Here, the expectation is taken with respect to the data symbols and the noise. Defining matrices $R$, $p$ and $N$ as

$$\mathbf{R} = E\left[\gamma[n]\gamma^T[n]\right] \tag{19}$$

$$\mathbf{p} = E\left[d(n)\gamma[n]\right] \tag{20}$$

$$\mathbf{N} = E\left[\eta[n]\eta^T[n]\right] \tag{21}$$

The equalizer taps are given by

$$c = (R+N)^{-1}.p \tag{22}$$

and the MMSE is

$$J_{min} = \sigma_d^2 - p^T(R+N)^{-1}.p \tag{23}$$

$$\sigma_d^2 = E[|d(n)|^2]$$

Evaluating the expectation over $R$ and $p$ with respect to the data and the noise, we have

$$p = \left[\alpha_K ... \alpha_0 ... \alpha_{-K_2}\right]^T \quad R = \left[r_{i,j}\right]_{K_1+K_2+1,K_1+K_2+1} \tag{24}$$

Where

$$r_{i,j} = \phi^T F_{i,j}\phi \quad \text{and} \quad F_{i,j} = \left[f_{1,k}\right]_{n_1+n_2+1,n_1+n_2+1}$$

$$f_{lk} = \begin{cases} 1, & l-k = j-i \\ 0, & l-k \neq j-i \end{cases} \tag{25}$$

$$N = E\left[\eta[n]\eta^T[n]\right] = \frac{N_0}{2}.\left(\sum_{i=1}^{L}\beta_i^2\right).I_{K_1+K_2+1} \tag{26}$$

Where $I$ is the identity matrix. This Rake-equalizer receiver will eliminate ISI as far as the number of equalizer's taps gives the degree of freedom required. In general, the equalizer output can be expressed as

$$\hat{d}(n) = q_0.d(n) + \sum_{i=0} q_1.d(n-i) + w(n) \tag{27}$$

with $\qquad q_n = \alpha_n.c_n$

The variance of $w(n)$ is

$$\sigma_{w(n)}^2 = \left(\sum_{i=-K_1}^{K_1} c_i^2\right)\left(\sum_{l=1}^{L}\beta_l^2\right).E_p.\frac{N_0}{2} \tag{28}$$

Where $E_p$ is the pulse energy.

the case of DFE, assuming error free feedback, the input data vector can be written in the form of

$$\gamma_{DFE}[n] = [\Phi^T d[n+K_1] ... \Phi^T d[n] d[n-1] ... d_{n-K_2}] \tag{29}$$

Using the same approach as for the linear equalizer, the MMSE feedforward taps for tap equalizer are obtained as

$$C_{DFE} = (R_{DFE} + N_{DFE})^{-1} p_{DFE} \tag{30}$$

Where $\qquad C_{DFE} = [c_{-K1} .... c_0 \, 0 .... 0]$

Also $\qquad R_{DFE} = \begin{bmatrix} R_F & U^T \\ U & I_{K_2} \end{bmatrix} \tag{31}$

Where $R_f = (K_1 + 1)$ square matrix with its element defined by (19) .Matrix $U$ is defined by

$$U = [u_{ij}]_{i=1,\ldots,K_2}$$
$$= 1,\ldots,K_1+1 \tag{32}$$

Matrix $N_{DFE}$ and vector $p_{DFE}$ are given by

$$N_{DFE} = \begin{bmatrix} N_0/2 \sum_{l=1}^{L} \beta_l^2 I_{K_{l+1}} & O_{K_1+1,K_2} \\ O_{K_2,K_1+1} & O_{K_2,K_2} \end{bmatrix} \tag{33}$$

$$p_{DFE} = [\alpha K_1 \ldots \ldots \alpha_0 \; 0 \ldots \ldots 0]$$

Where matrix 0 is the all zero matrix.

The MMSE feedback taps are then obtained in terms of feed forward taps and matrix $U$.

$$[c_1 \ldots \ldots \ldots c_{k2}] = [c_{-K1} \ldots \ldots \ldots c_0] U^T \tag{34}$$

Conditioned on a particular channel realization, $h = [h_1 \ldots \ldots h_I]$,an upper bound for the probability of error using the chernoff bound technique given by

$$P(\hat{d}_n \neq d_n | h) \leq \exp\left(-\frac{1 - J_{\min}/\sigma_d^2}{2J_{\min}}\right) \tag{35}$$

An exact BER expression with independent noise and ISI terms can be expressed as a series expansion is given by

$$P(\hat{d}_n \neq d_n | h) = \frac{1}{2} - \frac{2}{\pi} \sum_{\substack{z=1 \\ z \, odd}}^{\infty} \frac{\exp(-z^2 w^2/2)\sin(zwq_0)}{z} \times \prod_{\substack{n=N_1 \\ n \neq 0}}^{N_2} \cos(zwq_n) \tag{36}$$

Note that ISI comes from the interfering symbols in the range of $N_1 T_s$ and $N_2 T_s$. Parameter z and w determine the accuracy of the error rate given by (35). Set the $q_i$' s that are within the span of the feedback taps to be 0, which corresponds to zero post-cursor ISI for the span of feedback taps.

## 5. SIMULATION STUDY AND ANALYSIS

### 5.1. Signal Waveform

The pulse shape adopted in the numerical calculations and simulations is the second derivative of the Gaussian pulse given by

$$w(t) = [1 - 4\pi(t/\varepsilon)^2] \exp(-2\pi(t/\varepsilon)^2) \tag{37}$$

### 5.2. Channel Model Parameter (IEEE 802.15.3a)

As we mentioned it before, we study the case of UWB channels CM3 and CM4 channel models. We have used an oversampling factor of eight for the root raised cosine (RRC) pulse. According to this sampling rate, time channel spread is chosen equal to 100 for CM4 and 70 for CM3, this corresponds to respectively 12 =100 / 8 and 9 = 70 / 8 transmitted symbols. This choice enables to gather 99% of the channel energy. The coherence bandwidths of CM3 and CM4 simulation are 10.6 MHz and 5.9 MHz respectively. The data rate is chosen to be 200 Mbps, one of the optional data rates proposed for IEEE standard. The size of the transmitted packets is equal to 2560 BPSK symbols including a training sequence of length 512. CIR remains constant over the time duration of a packet. The root raised cosine (RRC) pulse with roll off factor $\alpha$ =0.5 is employed as the pulse-shaping filter. The CM3 and CM4 indoor channel model is adopted in simulation. The simulated channel impulse responses for CM3 and CM4 are shown in figure 4 and figure 6. The power delay profiles for CM3 and CM4 are plotted in figure 5and figure 7 respectively: The simulation parameter settings for the two channel models are listed in Table.1. The pulse waveform and power spectral density are showed as figure 3.

Table.1 Parameter Settings for IEEE UWB Channel Models

| Scenario | $\Lambda$ (1/ns) | $\lambda$ (1/ns) | $\Gamma$ (1/ns) | $\gamma$ (1/ns) | $\sigma_\xi$ (dB) | $\sigma_\varsigma$ (dB) | $\sigma_g$ (dB) |
|----------|--------|--------|--------|--------|--------|--------|--------|
| CM3 | 0.0067 | 2.1 | 14 | 7.9 | 3.3941 | 3.3941 | 3 |
| CM4 | 0.0067 | 2.1 | 24 | 12 | 3.3941 | 3.3941 | 3 |

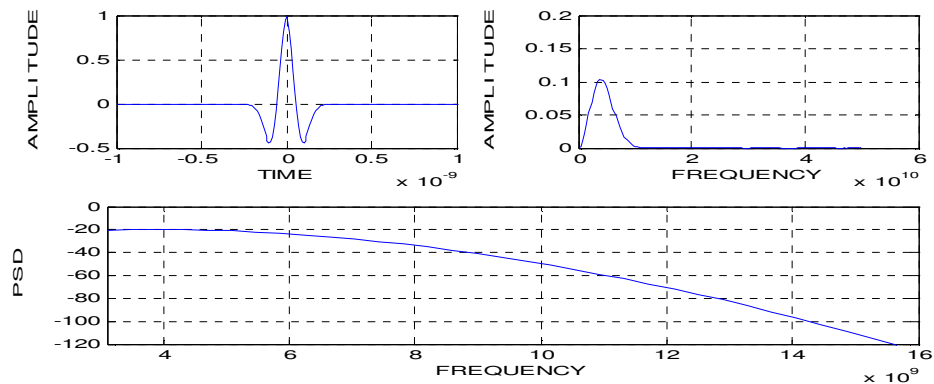

Figure.3. Second derivative of Gaussian pulse

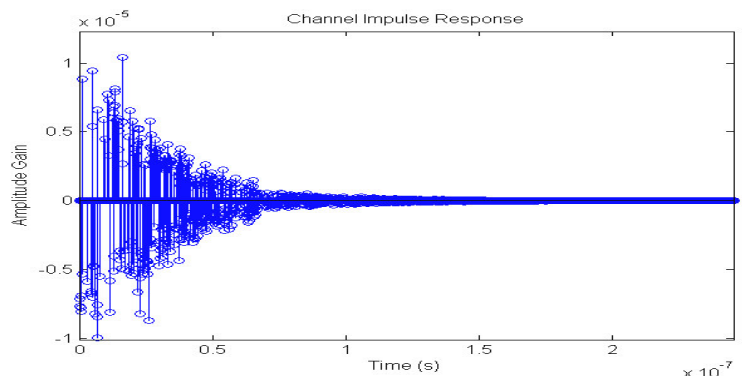

Figure.4. Channel Impulse Response of CM3 (NLOS)

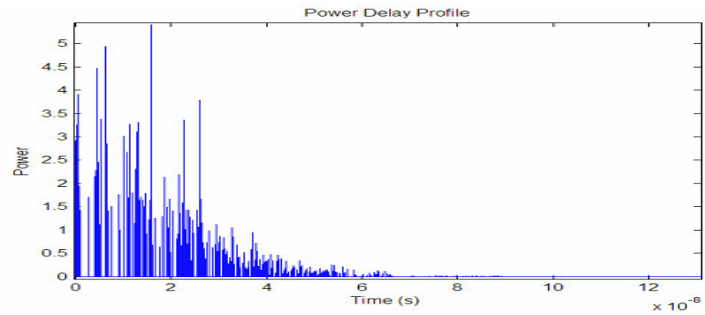

Figure.5. Power Delay Profile of CM3 Channel model

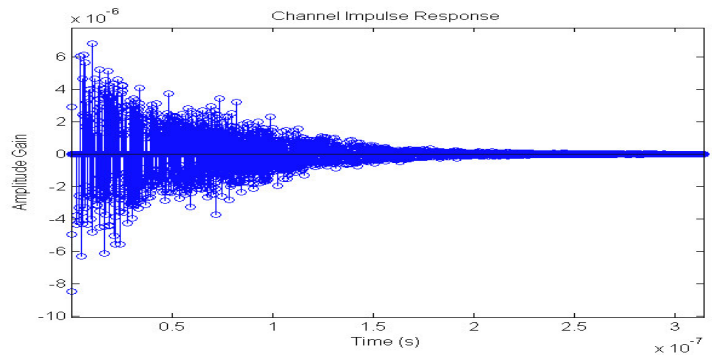

Figure.6. Channel Impulse Response of CM4 (NLOS)

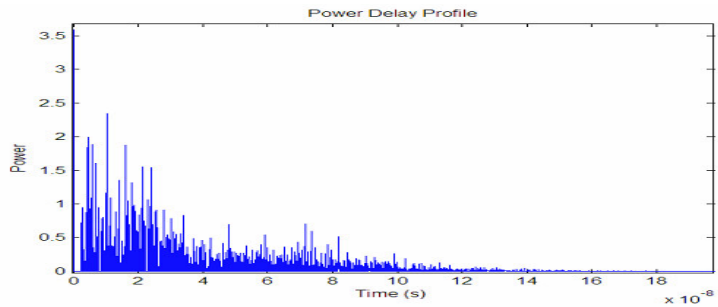

Figure.7. Power Delay Profile of CM4 Channel model

### 5.3. BER ANALYSIS

In the case of time domain equalization, we have at first to optimize the number of Rake fingers and the number of equalizer taps. The Rake fingers are regularly positioned according to time channel spread and the number of fingers. For example, in the case of CM4 channel, with $L = 10$. Figure.8 and figure.9 show the effect rake, MMSE and Rake-MMSE using Monte-Carlo simulation. The Rake combiner output at time $t = n.T_s$. The BER simulation results obtained using CM3 channel data is shown in figure.(8). As expected, using an MMSE equalizer to compensate for ISI, a relative improvement is observed. The major comparison lies in the Rake-MMSE receiver versus the Rake receiver. Rake-MMSE receiver has around1.8dB performance improvement compared to a channel with rake receivers for CM3 channel.

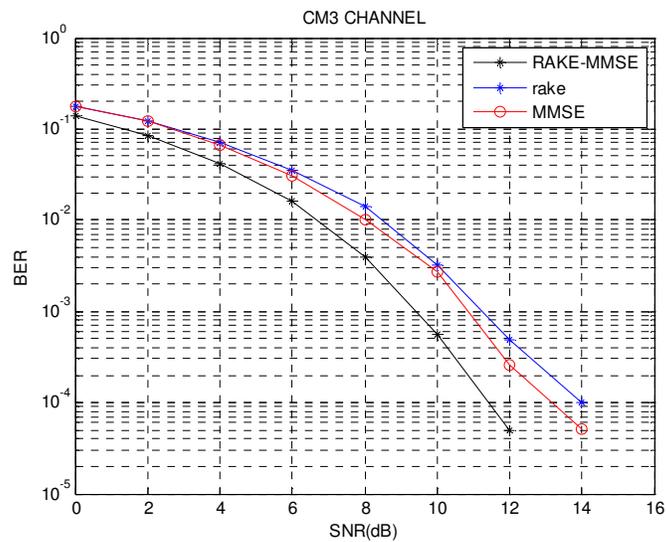

Figure.8. Performance of UWB receivers for CM3 channel model

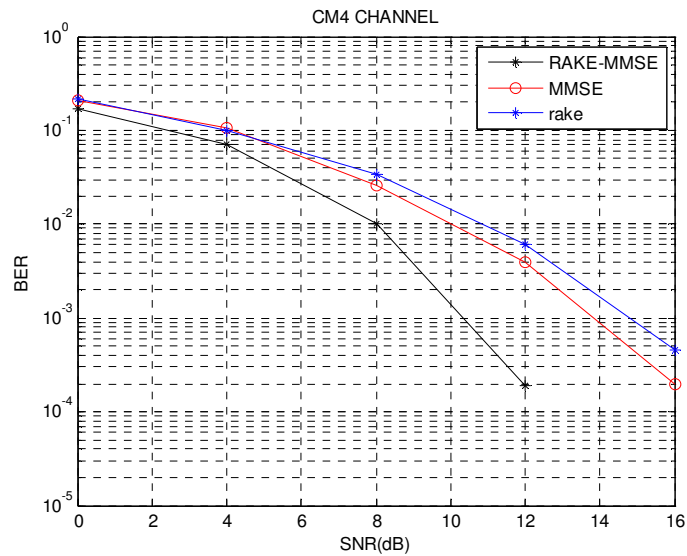

Figure.9. Performance of UWB receivers for CM4 channel model

Using CM4 channel data, the simulation results obtained as shown in figure.9.Using Rake-MMSE, i.e. Rake-MMSE and a rake receiver a gain of around 4dB is observed. This result can be explained by considering the fact that at high SNR's it is mainly the ISI that affects the system performance whereas at low SNR's the system noise is also a major contribution in system degradation (more signal energy capture is required). The performance dramatically improves when the number of Rake fingers and the equalizer taps are increased simultaneously in Rake-MMSE receiver. Performance of UWB Rake-MMSE-receiver for different number of equalizer taps and rake fingers for CM3 channel model as shown in figure.10. At $10^{-3}$ BER floor DFE provides more than 2dB SNR improvement than that of LE for CM3 channel model.

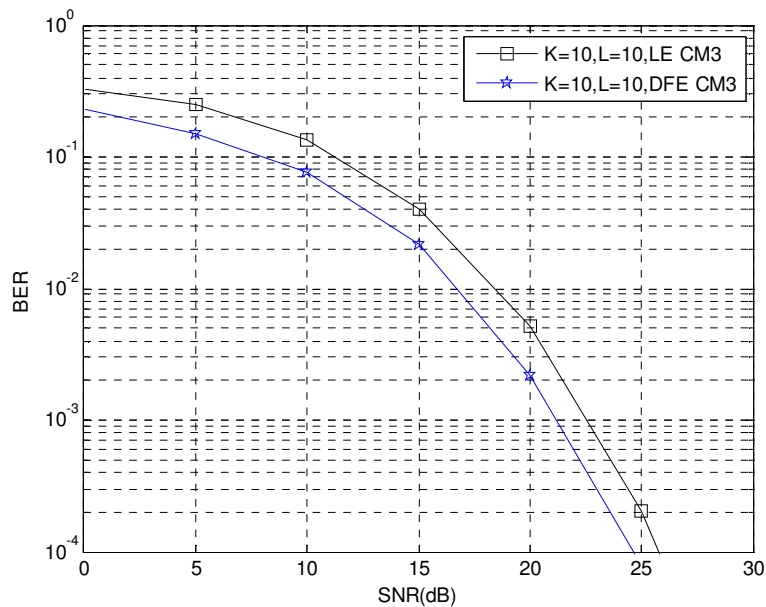

Figure.10. Performance of UWB rake-MMSE-receiver for different number of equalizer taps
and rake fingers for CM3 channel model

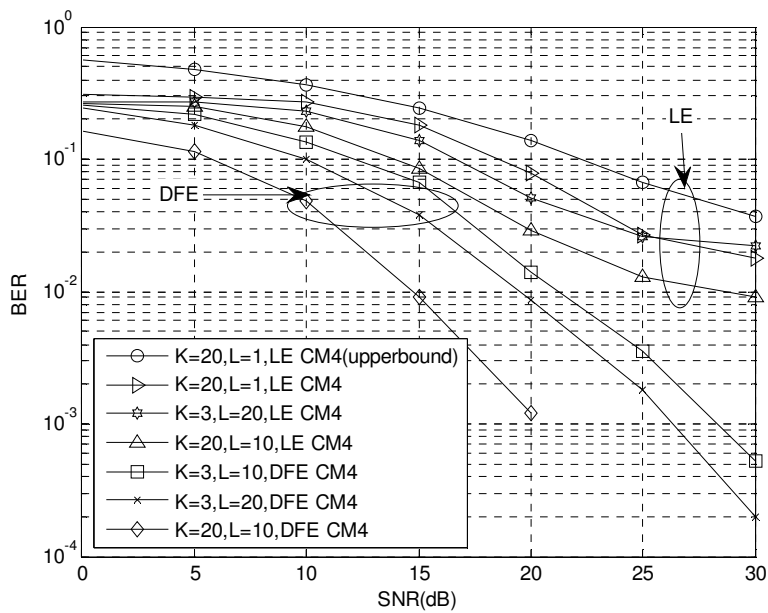

Figure.11. Performance of UWB rake-MMSE-receiver for different number of equalizer taps
and rake fingers for CM4 channel model

At low to medium SNR's, however, the receiver with more Rake fingers outperforms the one
that has more equalizer taps but fewer Rake fingers. This result can be explained by considering
the fact that at high SNR's it is mainly the ISI that affects the system performance whereas at
low SNR's the system noise is also a major contribution in system degradation (more signal
energy capture is required). The performance dramatically improves when the number of Rake fingers and the equalizer taps are increased simultaneously, i.e. K = 20, L = 10 as shown in figure.11. As expected the receiver has better performance over CM3 with smaller delay spread than CM4. Again BER performance observed on different UWB NLOS channel models (CM3 and CM4) shows that LE fails to perform satisfactorily at high SNR's due to presence of zeros outside the unit circle. These difficulty BER floor can be overcome DFE structure. A DFE outperforms a linear equalizer of the same filter length, and the performance further improve with increase in number of equalizer taps.DFE performances are computed by Monte-Carlo computer simulations, using a training sequence with length 500.

## 6. CONCLUSION

UWB is emerging as a solution for the IEEE 802.15a (TG3a) standard which is to provide a low complexity, low cost, low power consumption and high data-rate among Wireless Personal Area Network (WPAN) devices. For high data rate short range, the receivers combats inter-symbol interference by taking advantage of the Rake and MMSE equalizer structure by using different UWB channel models CM3 and CM4. For a MMSE equalizer operating at low to medium SNR's, the number of Rake fingers is the dominant factor to improve system performance, while, at high SNR's the number of equalizer taps plays a more significant role in reducing error rates. One can observe a BER floor at high SNR's the receiver has better performance over CM3 with smaller delay spread thanCM4 in non-line of sight indoor channel models . Rake-MMSE receiver has around1.8dB performance improvement compared to a channel with rake receivers for CM3 channel model. Again BER performance observed on different UWB channel models (CM3 and CM4) shows that DFE outperforms a LE of the same filter length, and the performance further improve with increase in number of equalizer taps. It is concluded that increasing the number of rake fingers performance become superior at low to medium SNR. These architecture has opened up new directions in designing efficient time domain equalizers for UWB system and can be implemented in DSP processors for real - time applications.

**Authors**

Bikramaditya Das received his B.Tech in Electronics and Telecommunications Engineering from the University of B.P.U.T, Rourkela, Orissa, India, in 2007. From 2008 to till now he is a research Fellow under the Department of Electrical Engineering at the N.I.T, Rourkela, India. He is a Member IEI. 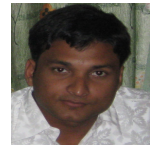

Dr.Susmita Das, Ph.D, is Associate Professor of Electrical Engineering, NIT, and Rourkela, India. She has twenty years of teaching and research experience and has many research papers to her credit. She is Member IEEE, Fellow IETE, LM ISTE and Member IEI. Her research interests include Wireless Communication, DSP, and Application of Soft Computing Techniques etc. 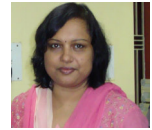